\pdfoutput=1
\documentclass[preprint,aps]{revtex4}
\usepackage{epsfig}
\usepackage{graphics}
\usepackage{graphicx}

\begin{document}

\title{\bf Some Radiative Corrections to Neutrino Scattering: \\
I Neutral Currents}
\author{James P.\ Jenkins}\email
{ jjenkins6@lanl.gov} 
\author{T.\ Goldman}\email{ tgoldman@lanl.gov}
\affiliation{Theoretical Division,  
Los Alamos National Laboratory, Los Alamos, NM 87545}

\begin{flushright}
June 4, 2009\\
{LA-UR-09-02131}\\
{arXiv:yymm.nnnn}\\
\end{flushright}

\begin{abstract}

With the advent of high precision neutrino scattering experiments 
comes the need for improved radiative corrections. We present a 
phenomenological analysis of some contributions to the production 
of photons in neutrino neutral current scattering that are relevant 
to experiments subsuming the 1\% level.

\end{abstract} 


\maketitle

\section{Introduction}\label{sec:Introduction}

Recent neutrino scattering experiments, particularly those searching 
for electron neutrino appearance~\cite{LSND,MB_NuRes,MB_NuBarRes} by oscillation from 
muon neutrino beams, report signals below the 1\% level. Such 
unprecedented levels of sensitivity demand corresponding efforts to 
determine backgrounds. Radiative corrections are clearly expected 
at this level. 

Many years ago, phenomenological analyses of radiative corrections 
were applied to studies of pion decay~\cite{MarcianoSirlin,GoldmanWIlson}, 
to constrain corrections to the weak interaction from non-V-A 
currents. Their accuracy was confirmed by precise calculations when 
the Standard Model became established. 

Although many contributions are possible, obviating unambiguous 
predictions, similar phenomenological analyses can provide overall 
scale, energy and angular distributions for photon production in 
neutrino scattering.  Such methods were used in~\cite{Choban:1977hf,Gershtein:1980wu} 
to help explain the Gargamelle single photon signal~\cite{Gargamelle}. 
Armed with these predictions, experimental results can then be used 
to determine whether the total of such contributions are significant 
within each experiment. 

Of course, Monte Carlo calculations by experimental groups include bremsstrahlung 
from charged particles in the final state and radiative decays of $N^{*}$
and $\Delta$ resonances. However, these are $s$-channel effects. We 
consider here photons produced from the particles being exchanged in 
the $t$-channel, and so less obviously connected to the external line 
quanta. In general, such contributions are necessary simply to maintain gauge invariance. 

We provide here two examples of amplitudes relevant to neutral current 
neutrino scattering, which are especially important when distinguishing final state electrons 
and photons is difficult. We eschew details such as 
interferences between different contributions, as the final state distributions 
in our examples differ significantly, minimizing the overlap. (In addition, 
it is usually the case that possible interfering contributions involve the 
exchange of much higher mass particles with much smaller phenomenological 
couplings~\cite{GG}; hence, interference effects are expected to be minimal.) Our goal is 
simply to provide experimental groups with distribution formulas that 
they can use to extract these contributions or determine limits on them. 

We provide results for both neutrino and antineutrino scattering. Our 
results may be easily extended to other similar contributions both in 
neutral and charged current scattering. Although we focus on modest 
energy studies, our results are fully relativistically covariant and thus may 
be applied to any energy. Additional contributions or Regge trajectory 
generalizations of the meson exchanges are likely necessary for application 
to very high energies.  Our results are also straightforwardly generalizable to other related 
processes such as charged current interactions, which we will present 
elsewhere. 

\section{Lagrangian}\label{sec:Lagrangian}

Photon production in neutral current neutrino scattering produces a 
background to the identification of electron neutrino events in (generally 
non-magnetic) detectors which distinguish poorly between electrons 
and photons.  Standard radiative corrections, such as those due to 
bremsstrahlung of a photon from charged particles in the target or final 
state, have been closely examined previously~\cite{CoherentProdPhotonByNu}.  
See also~\cite{GEANT4,nuance} and references therein.
Additional 
processes, such as production of a $\Delta$ baryon or N$^{*}$ followed 
by its decay back to a nucleon and a photon~\cite{PhoteElPiResProd}, are also already included 
in many experimental Monte Carlo assessments of backgrounds~\cite{GG}. 

Recently, however, a ``new'' triangle anomaly has been identified in 
reference \cite{H3}, hereinafter referred to as H3, which can contribute a 
previously unconsidered source of such background. However, the 
structure of that contribution is very similar to a phenomenological 
contribution which we display in Fig.(\ref{fig:NuNGammaDiag}). Here 
the Z-boson carrying the neutral current interaction from the neutrino line 
mixes into a vector boson, in the familiar fashion of Vector Meson Dominance 
(VMD).~\cite{vmd}  In this case, the hadronic vector meson is an $\omega$ 
(or a $\rho^{0}$) which undergoes a virtual decay to the photon of interest 
and a pion in the $t$-channel.  This last couples strongly to the hadron target 
(nucleon or nucleus) completing the interaction.

The advantage over the H3 approach is that the vertex strength is known 
phenomenologically from the rate of the decay
$$
\omega \rightarrow \pi^{0} + \gamma 
$$
and similarly for the $\rho^{0}$ case. The pion coupling to the nucleon 
is also well-known~\cite{LatticePiNNFormFactor}.  The strength of the $Z-\omega$ 
mixing is determined by VMD and the off-shell variation of this mixing is easily 
determined by a well-known extension~\cite{GHT1,GHT2}, originally worked out 
for the case of isospin violating contributions to the nuclear force. We explicitly calculate 
the appropriate analog in Appendix~(\ref{subsec:XomegaMixing}) for completeness. 

No other parameters are required, so the prediction of the contribution to 
the total cross section for producing a final state photon is absolute for 
this graph.  We emphasize that the structure of the central vertex has the 
same vector-vector-axial-vector coupling structure as in H3, due to the 
axial vector nature of the pion current. However, in our approach, all of 
the strong interaction corrections (higher order in QCD and quark-antiquark 
``resonance'' effects) relevant to the triangle graph of H3 are fully taken into 
account phenomenologically. Of course, other similar contributions occur 
with vector-meson recurrences, etc., but these predominantly affect only the 
overall strength, and furthermore, experience has shown that at the modest 
energies of the experiments of interest (LSND~\cite{LSND} and MiniBooNE~\cite{MB_NuRes,MB_NuBarRes}), 
the sum over all such contributions is likely to be dominated by these leading 
ones. 

The interaction Lagrangian terms needed in our approach are: 
\begin{equation}
{\cal L}_{I} = e g_{\omega\gamma\pi} \epsilon_{\mu\nu\xi \sigma} 
\omega^{\mu}\partial^{\nu}\pi^{0}F^{\xi \sigma} 
+ g_{\pi NN} \bar{\Psi}\gamma^{\mu}\partial_{\mu}\vec{\pi}\cdot \vec{\tau}\Psi
\end{equation}
where $\Psi$ is the nucleon field and $ \vec{\tau}$ is the vector of usual isospin 
generators. 

We also make use of the usual Standard Model weak interaction couplings 
of the $Z$-boson to neutrinos and to quarks. The last are needed to compute the 
off-shell variation of the VMD mixing between the $Z$ and the $\omega$. The 
${\omega-\pi-\gamma}$ coupling constant,  $g_{\omega\pi\gamma}=e g_\omega$ (with the 
explicit factor of the electromagnetic coupling, $e$, stripped out for clarity) is 
determined from the experimental value of the (on-shell) radiative $\omega 
\rightarrow \pi + \gamma$ decay rate, shown  in Appendix~(\ref{subsec:omegaDecay}). 
We will discuss possible off-shell effects on this compared with the corresponding 
effects for the H3 approach in our discussion section (\ref{sec:Discussion}).


\section{Cross Section}\label{sec:CrossSection}
\begin{figure}
\includegraphics[scale=0.65]{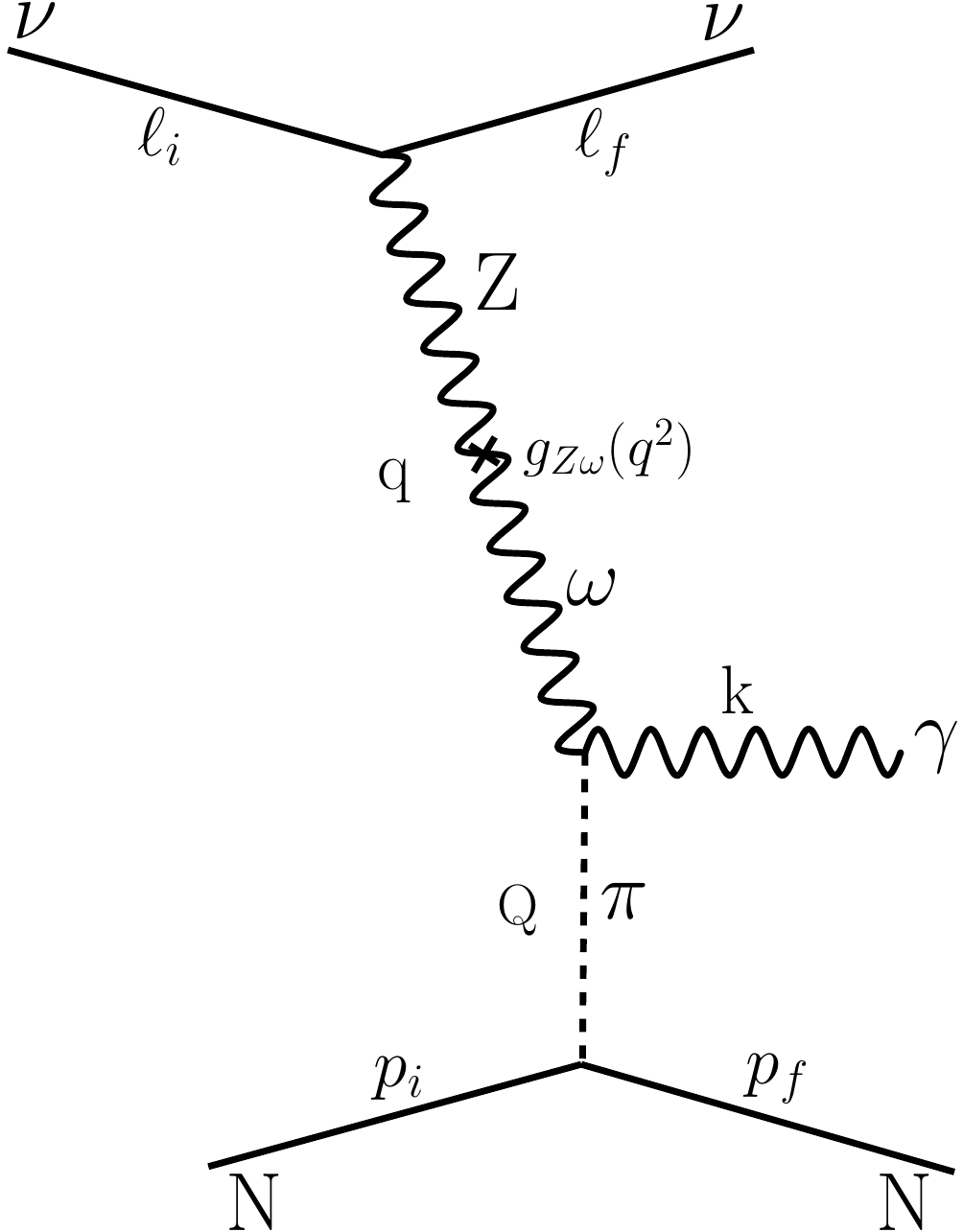}
\caption{Specific diagram considered in this analysis.  Variants are discussed 
in the text.}
\label{fig:NuNGammaDiag}
\end{figure}

We consider the process shown in Fig.\ref{fig:NuNGammaDiag}.  In terms of 
the labeled momentum 4-vectors, the squared amplitude is
\begin{equation}
 \mathcal{A}^2 = \frac{128 M_n^2 g_{\nu Z}^2 g_{\omega\gamma\pi}^2 g_{\pi N N}^2 
 g_{\omega Z}^2(q^2)}{(q^2-M_Z^2)^2(q^2 - M_\omega^2)^2(Q^2-M_\pi^2)^2} \ell_i 
 \cdot \ell_f (p_i \cdot p_f-M_N^2)\left((k \cdot \ell_i)^2 + (k \cdot \ell_f)^2 \right),
\end{equation}
where the neutrino-$Z$ coupling is given by $g_{\nu Z} = (g/2)\cos\theta_W$ in terms 
of the $SU(2)$ gauge coupling $g$.  The remaining coupling constants will be discussed 
in what follows.  In the center of mass (CM) frame, the momenta can be written explicitly as
\begin{eqnarray}
 \ell_i &=& (E_{\ell_i},\vec{E}_{\ell_i}) \\
p_i &=& (E_{p_i},-\vec{E}_{\ell_i}) \\
 \ell_f &=& (E_{\ell_f},\vec{E}_{\ell_f}) \\
p_f &=& (E_{p_f},\vec{p}_{p_f}) \\
 k &=& (E_{k},\vec{E}_{k}),
\end{eqnarray}
where we employ the shorthand $\vec{E}_{i} = \vec{p}_{i}$ to indicate a massless particle's 3-momentum.  To obtain the desired cross-section, we first partially evaluate the phase space integrals 
by making use of momentum conserving delta functions.  We find the differential cross-section 
for the photon in the final state is
\begin{equation}
\label{eq:diffsigma}
 \frac{d\sigma}{dE_kd\mu} = \frac{M_N^2g_{\nu Z}^2 g_{\omega\gamma\pi}^2 g_{\pi N N}^2}
 {(2\pi)^4 E_{\ell_i}(E_{\ell_i} + E_{p_i})}\int dE_{\ell_f} d \phi \frac{g_{\omega Z}^2(q^2) q^2 
 Q^2 \left((k \cdot \ell_i)^2 + (k \cdot \ell_f)^2 \right)}{(q^2-M_Z^2)^2(q^2-M_\omega^2)^2 
 (Q^2 - M_\pi)^2},
\end{equation}
in terms of its energy ($E_k$) and opening angle from the beam direction ($\mu = \cos\theta$).  
Here the $t$-channel momentum transfers are: $q^2 = -2E_{\ell_i}E_{\ell_f}(1-\mu_{\ell_f})$ and 
$Q^2 = q^2 - 2k \cdot \ell_i + 2k \cdot \ell_f$, where $\mu_{\ell_f}$ is the cosine of the opening 
angle between the neutrino in the final and initial state.  It is related to $\mu$ and the cosine 
 $\mu_{\ell_f k}$ of the opening angle between the photon and the final state neutrino by 
\begin{equation}
\mu_{\ell_f} = \mu\mu_{\ell_f k} + \sqrt{1-\mu^2}\sqrt{1 - \mu_{\ell_f k}^2}\cos\phi.
\end{equation}
This is the only source of $\phi$ dependence in the system.  Momentum conservation fixes 
\begin{equation}
\mu_{\ell_f k} = \frac{1}{2E_{\ell_f}E_k}\left( \sqrt{s}^2 - 2\sqrt{s}(E_{\ell_f} + E_{k}) + 
2E_kE_{\ell_f} - M_N^2 \right)
\end{equation}
as a function of the invariant $\sqrt{s} = E_{\ell_i} + E_{p_i}$.  Requiring that $\mu_{\ell_f k}
 \in \{-1,1\}$ yields the $E_{\ell_f}$ integration limits
\begin{eqnarray}
 E_{\ell_f}^{\rm min} &=& \frac{\sqrt{s}^2 - M_N^2 - 2\sqrt{s}E_K}{2\sqrt{s}}\\
 E_{\ell_f}^{\rm max} &=& \frac{\sqrt{s}^2 - M_N^2 - 2\sqrt{s}E_K}{2\sqrt{s} - 4E_k}.
\end{eqnarray}
Additionally, we require $E_k \leq \frac{1}{2\sqrt{s}}(\sqrt{s}^2 - M_N^2)$ to maintain positive 
values for these energies.  The $\omega - Z$ mixing coupling function, $g_{\omega Z}^2(q^2)$, 
is found from the self energy diagram in Fig.\ref{fig:ZOmegaMixing} of Appendix~(\ref{subsec:XomegaMixing}).  
It is almost constant in the space-like regime where it is needed here.

Taking the limit $|q^2| \ll M_Z^2$ and $|Q^2| \gg M_\pi^2$, the cross section simplifies to 
\begin{equation}
  \frac{d\sigma}{dE_kd\mu} = \frac{M_N^2g_{\nu Z}^2\bar{g}_{\omega Z}^2 g_{\omega\gamma\pi}^2 
  g_{\pi N N}^2}{(2\pi)^4 E_{\ell_i}(E_{\ell_i} + E_{p_i}) M_Z^4}\int dE_{\ell_f}\left((k \cdot \ell_i)^2 + 
  (k \cdot \ell_f)^2 \right)\int d\phi \frac{q^2}{Q^2(q^2-M_\omega^2)^2},
\end{equation}
where we have neglected the $q^2$ dependence of $g_{\omega Z}$ and assume an average value.  Integrating over 
$\phi$, we obtain
\begin{eqnarray}
 \frac{d\sigma}{dE_kd\mu} &=& \frac{M_N^2 E_k^2 g_{\nu Z}^2\bar{g}_{\omega Z}^2 
 g_{\omega\gamma\pi}^2 g_{\pi N N}^2}{(2\pi)^3 E_{\ell_i}(E_{\ell_i} + E_{p_i}) 
 M_Z^4} \int dE_{\ell_f} \left( E_{\ell_i}^2(1-\mu)^2 + E_{\ell_f}^2(1-\mu_{\ell_fk})^2 
 \right) \\ \nonumber
 &\times& \frac{1}{f^2(b-c)^2} \left\{ \frac{a-b}{(b^2-1)^{\frac{1}{2}}} + \frac{c^3 - 2ac^2 
 + abc - b + a}{(c^2 - 1)^{\frac{3}{2}}} \right\},
\end{eqnarray}
where we have introduced
\begin{eqnarray}
 f &=& 2E_{\ell_i}E_{\ell_f}\sqrt{1-\mu^2}\sqrt{1-\mu_{\ell_fk}^2}\\
 a &=& \frac{2E_{\ell_i}E_{\ell_f}(1-\mu\mu_{\ell_fk})}{f}\\
 b &=& \frac{2E_{\ell_i}E_{\ell_f}(1-\mu\mu_{\ell_fk}) + 2E_kE_{\ell_i}(1-\mu) - 2E_k
 		E_{\ell_f}(1-\mu_{\ell_fk})}{f}\\
 c &=& \frac{2E_{\ell_i}E_{\ell_f}(1-\mu\mu_{\ell_fk}) + M_\omega^2}{f}
\end{eqnarray}
to simplify the notation. The dimensionless quantities $a,~b$ and $c$ are all greater than 
unity, assuming physical parameter values.  This leaves only the one-dimensional $E_{\ell_f}$ 
integral to perform.

\subsection{Cross Section Characteristics}\label{subsec:character}

\begin{figure}
\includegraphics[bb=250bp 100bp 1427bp 811bp,clip,scale=0.39]{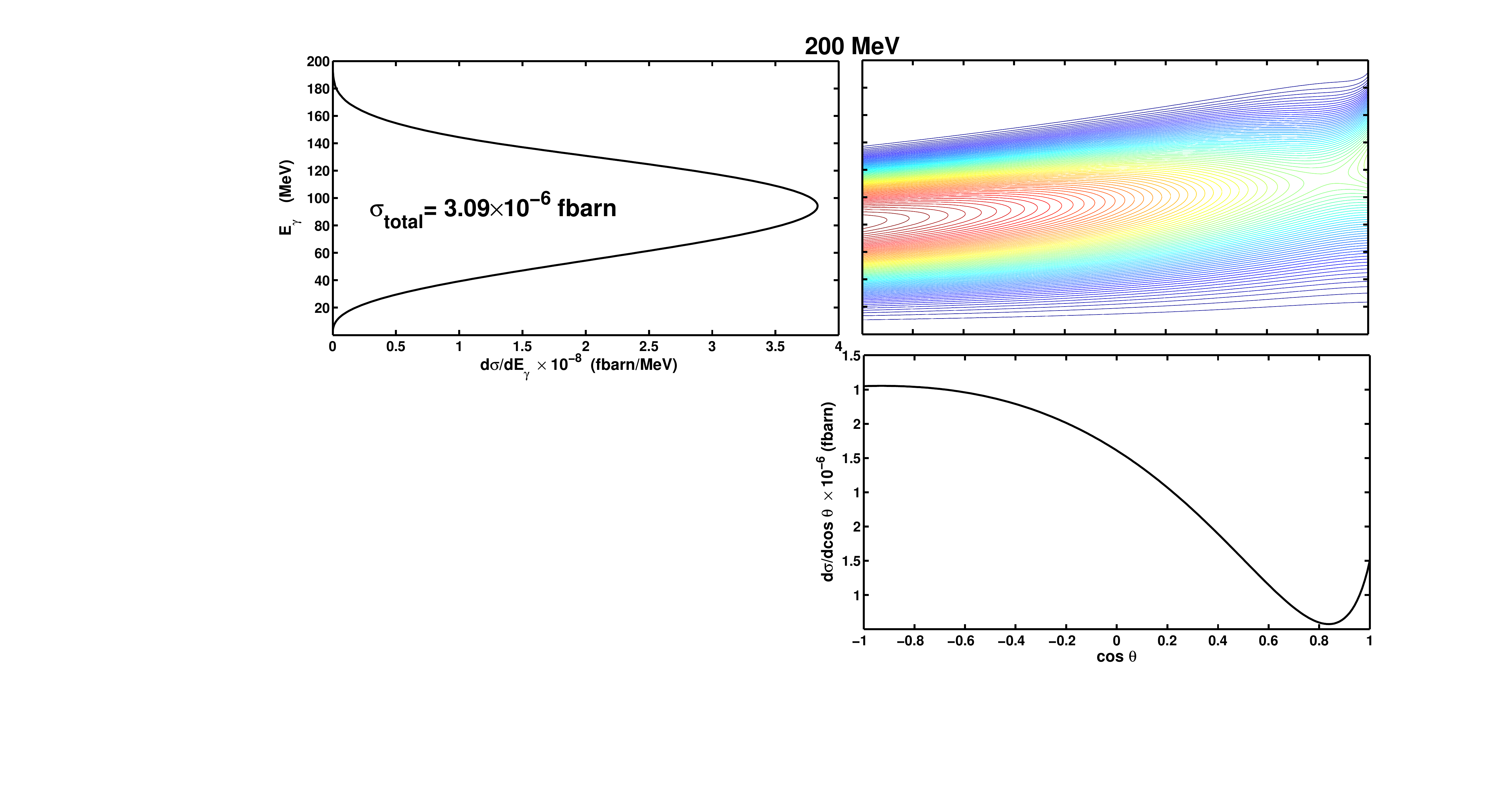}
\caption{Lab frame cross section with beam energy $E_{\ell_i} = 200~{\rm MeV}$.}
\label{fig:CS200MeV}
\end{figure}

\begin{figure}
\includegraphics[bb=250bp 100bp 1427bp 811bp,clip,scale=0.39]{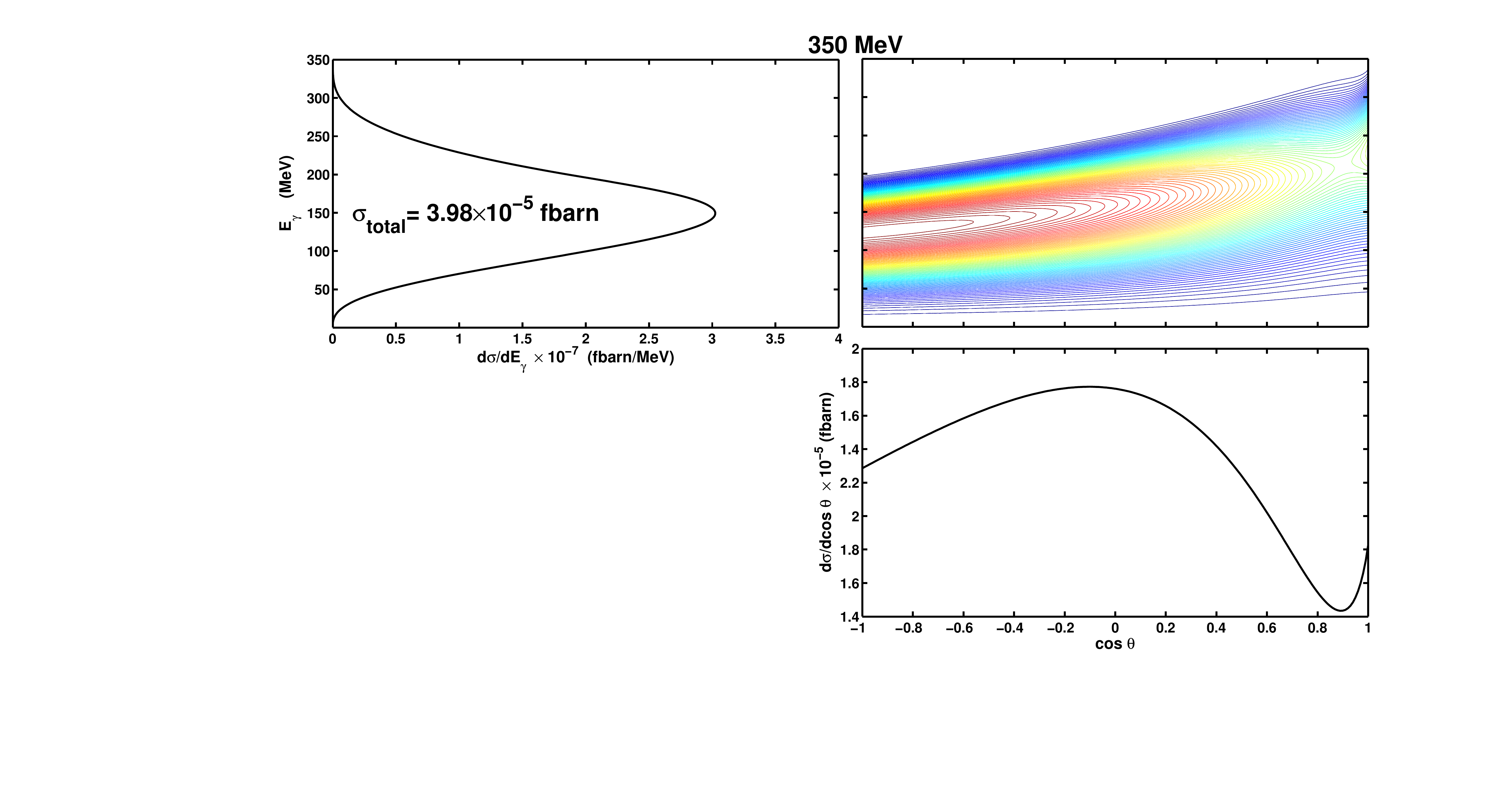}
\caption{Lab frame cross section with beam energy $E_{\ell_i} = 350~{\rm MeV}$.}
\label{fig:CS350MeV}
\end{figure}

\begin{figure}
\includegraphics[bb=250bp 100bp 1427bp 811bp,clip,scale=0.39]{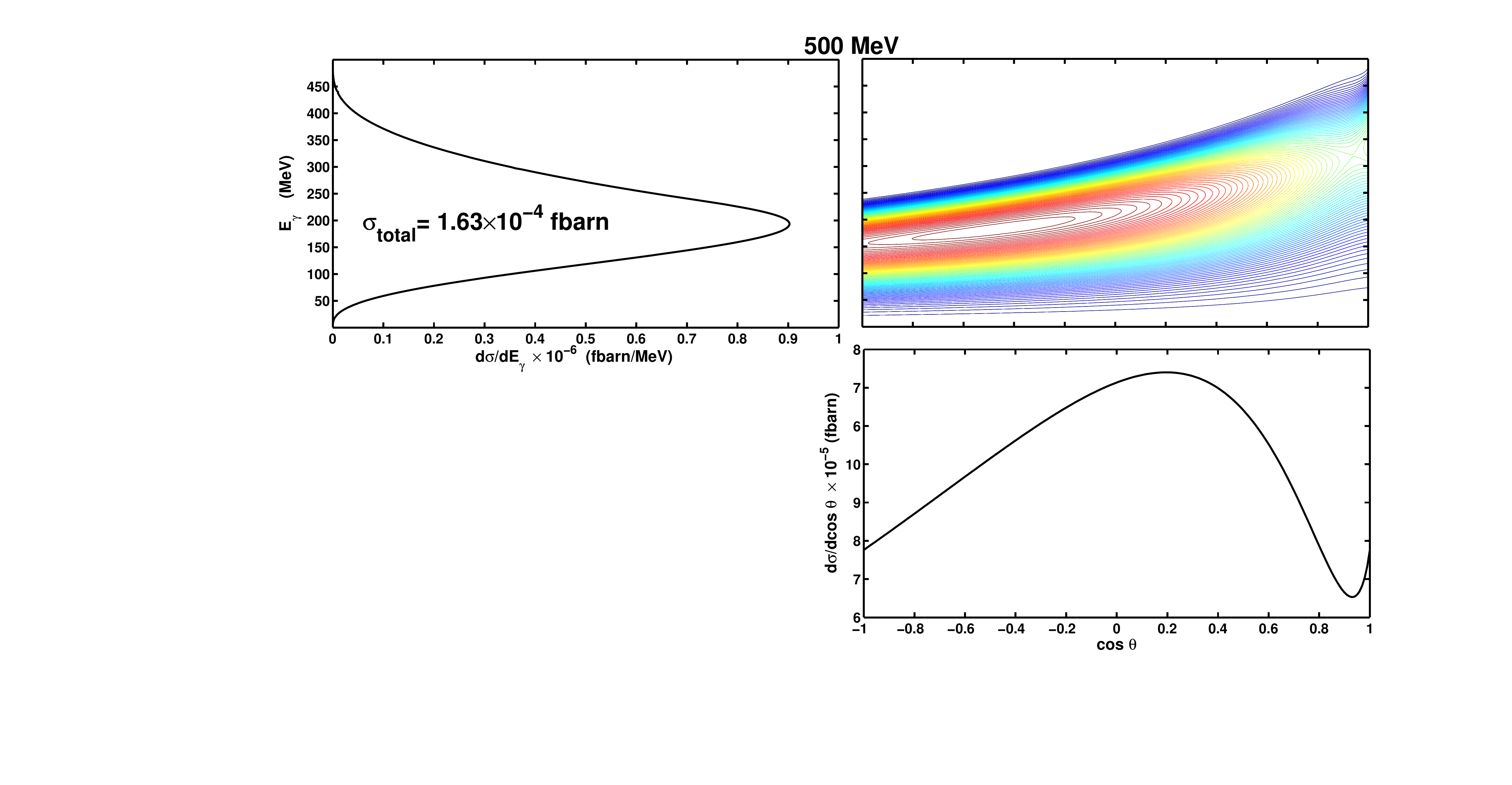}
\caption{Lab frame cross section with beam energy $E_{\ell_i} = 500~{\rm MeV}$.}
\label{fig:CS500MeV}
\end{figure}

\begin{figure}
\includegraphics[bb=250bp 100bp 1427bp 811bp,clip,scale=0.39]{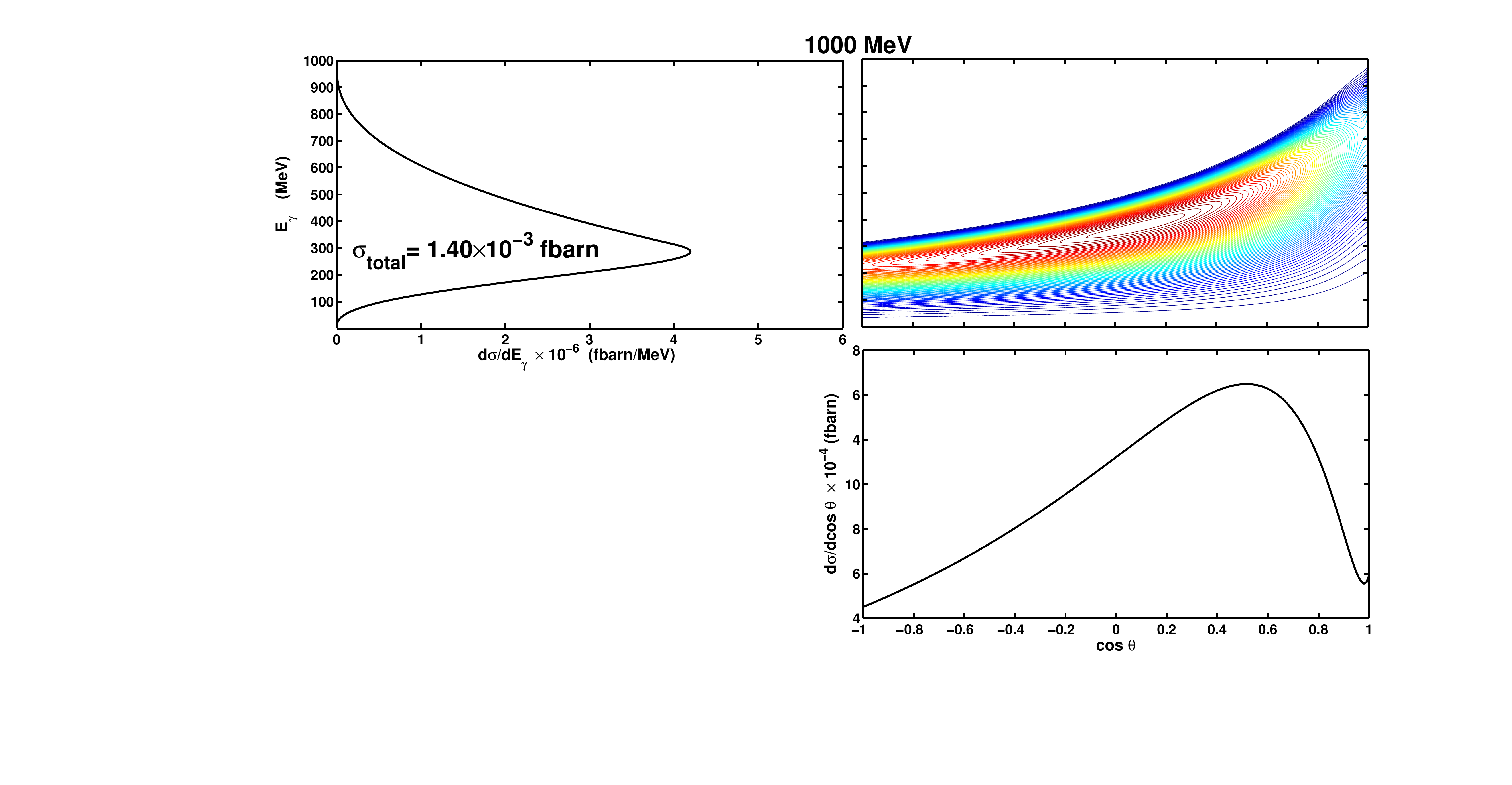}
\caption{Lab frame cross section with beam energy $E_{\ell_i} = 1000~{\rm MeV}$.}
\label{fig:CS1000MeV}
\end{figure}

In the CM form, the predicted cross section is weakly peaked in the backward direction 
with an energy maximum near the highest allowed energies due to the overall factor of 
$E_k$.  Boosting to the lab frame pushes the angular distribution forward and spreads 
the energy of the photon in the standard way. Numerically integrating Eq.~(\ref{eq:diffsigma}), 
we plot the differential cross-section for beam energies of $200~{\rm MeV}$, 
$350~{\rm MeV}$, $500~{\rm MeV}$ and $1000~{\rm MeV}$ in Figs.\ref{fig:CS200MeV}, 
\ref{fig:CS350MeV}, \ref{fig:CS500MeV} and \ref{fig:CS1000MeV}, respectively.  
These are boosted to the lab frame for convenience.  In each case, we display the 
$E_\gamma$ and $\cos\theta$ dependent contour plots as well as energy projection 
panels for the result of integrating over one of the variables.  The total cross section 
is also noted for reference.  The angular distribution moves toward the forward peak 
with increasing neutrino energy due to the growing boosts from the CM to the lab frame.  
The distribution consistently peaks near the center of the kinematically allowed 
photon energy range.

\begin{figure}
\includegraphics[scale=0.34]{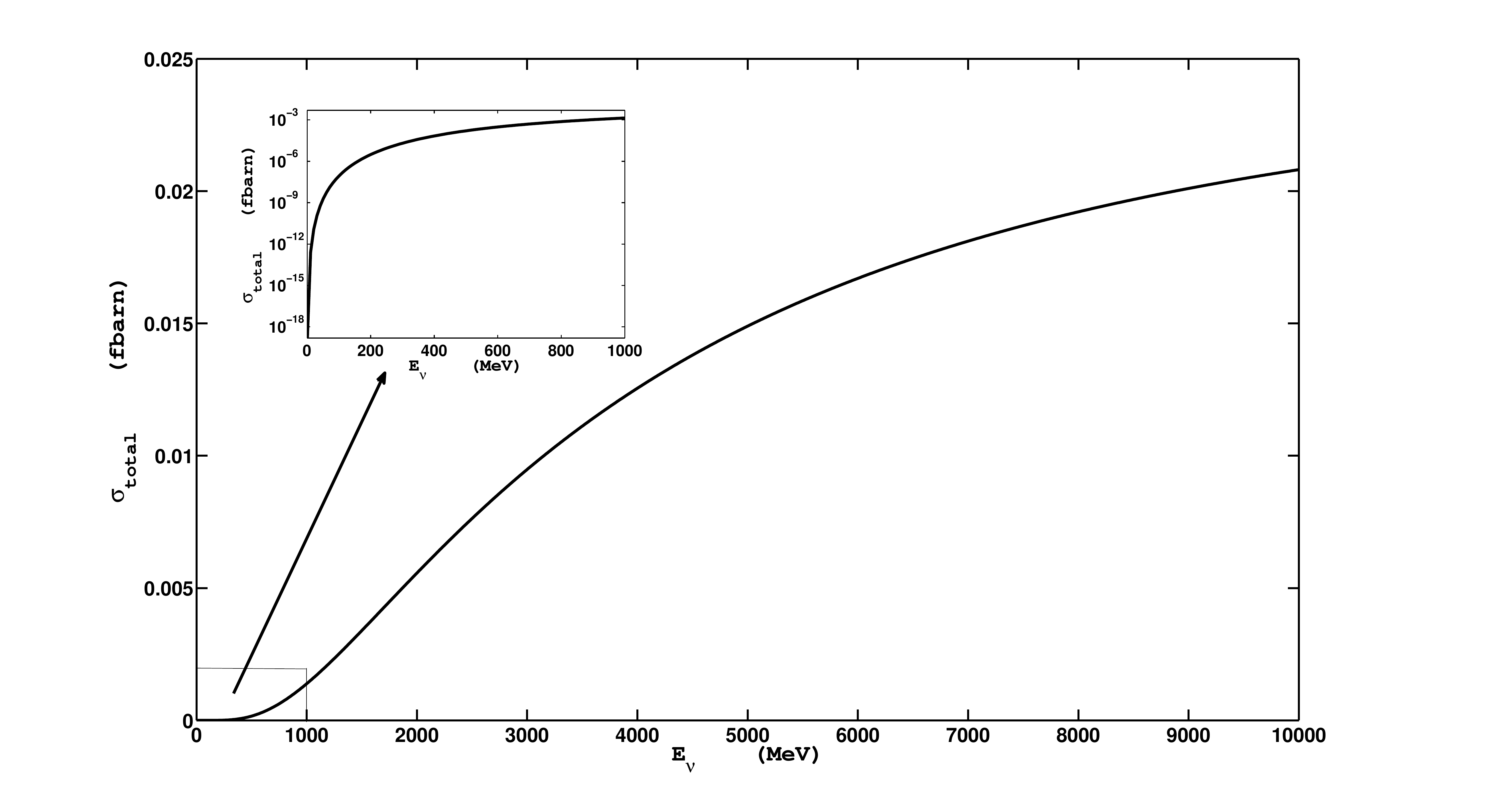}
\caption{Total cross section as a function of neutrino beam energy}\label{fig:SigTotMeV}
\end{figure}

Integrating over the final state photon energy and angular distribution, we plot the total 
cross section as a function of neutrino beam energy in Fig.\ref{fig:SigTotMeV}.  At 
high energies ($E_\nu \gg M_N$) the cross section grows as $\sqrt{E_\nu}$.  Near 
threshold, it grows as $E_\nu^2$, as can be seen in the logarithmic low energy insert plot.  
In the energy region below $1000~\rm{MeV}$, the cross section does not exceed 
$10^{-3}~\rm{fbarns}$, which is roughly three orders of magnitude less than the typical 
charged current cross sections at these energies~\cite{Lipari:1994pz}. Nevertheless, it is still at a 
level where it can affect current~\cite{MB_NuRes,MB_NuBarRes} and next 
generation~\cite{MicroBooneProposal} experiments that probe neutrino interactions with 
sub-percent sensitivities.  Current and proposed long baseline oscillation experiments 
(see, for example~\cite{MINOS_Res08,NOvAProposal,T2KProspects,BNL_NuWorkGroup} and 
references therein) will be sensitive to this class of processes with the order of magnitude 
enhanced cross sections shown in the high energy region of Fig.\ref{fig:SigTotMeV}. 


 \section{Discussion}\label{sec:Discussion}

Other processes, related to that in Fig.\ref{fig:NuNGammaDiag}, such as by exchange 
of the $\omega$ and $\pi^0$ for other mesons (with the same quantum numbers), will 
contribute to similar production of photons in the final state.  These will differ from our 
calculation here only by the meson masses and coupling constants.  

We also observe that this process yields identical results for both neutrino and anti-neutrino 
scattering.  The only difference between these resides in a sign change between the axial-vector 
neutrino-Z coupling.  The asymmetry term vanishes when Lorentz contracted with the rest of the 
diagram, which is symmetric under the free indices of the vertex.  (We have checked this feature 
by direct calculation.)  This is not the case for non-radiative neutral current neutrino scattering, 
where there is a difference between neutrino and anti-neutrino amplitudes.  Additionally, we find 
by means of direct computation that many variants of Fig.\ref{fig:NuNGammaDiag} vanish due 
to similar symmetry reasons. In particular,  
\begin{itemize}
 \item Amplitudes from diagrams with axial vector mesons replacing the vector 
 		$\omega$ vanish.
 \item The amplitude of the process with the $\pi$ and $\omega$ mesons interchanged 
 		is zero.
 \item Amplitudes from ``reversed'' diagrams (where the $\pi_0$ couples to the neutrino) 
 		vanish.
\end{itemize}  

We have used a phenomenological on-shell coupling strength for the $\omega-
\pi-\gamma$ vertex, which is a commonly used phenomenological approach. 
The slow off-shell variation of the computed $Z-\omega$ mixing supports such 
an approach, although a three-particle vertex may well behave differently. The 
variation one expects is for a decrease in amplitude as vertex form factors act 
to suppress the effective coupling~\cite{offsh}. This may be 
ultimately recovered by summing over higher mass Regge recurrences in an extension 
of our approach, which would eventually transition at very high energies to a purely 
quark picture of the couplings. 

Such a view raises the question as to whether the H3 axial anomaly 
might not give a more reliable evaluation of the strength of these processes after 
all. However, the use of the anomaly is only guaranteed accurate when currents 
coupling to the vertices are point-like. In the case of the pion, the axial current 
satisfies this condition, but for the $\omega$, being a composite object, the vector 
isoscalar current cannot provide the same guarantee. The vertex structure for 
$\omega-q-\bar{q}$ coupling produces the same kind of uncertainties in the H3 
approach as we encounter here; perhaps more so as the $\omega$ momentum 
wavefunction in the quark basis may only be known via Lattice QCD calculations, 
although crude quark model calculations do produce a 3-$\pi$ hadronic width consistent 
with experiment.~\cite{darw} We believe the issue is less serious 
for our calculation as the $Z-\omega$ mixing is effectively used only within a region of 
order only a few mass squared units from the on-shell point. 

Our view is that for low to modest energy scattering, such as is relevant to the 
experiments discussed here, our phenomenological approach is as valid as any 
other, although it is most reliable for momentum transfer and angular distributions, 
and least reliable for absolute rates. The most certain point is that these $t$-channel 
processes will be the same in neutrino and in anti-neutrino scattering, so a 
comparison of the two running modes for MiniBooNE will set the most stringent 
limits on the strength of these contributions to the total rate. 

Another significant background also occurs when a neutral pion is produced 
instead of a photon, but one of the photons from the decay of this pion is lost to 
the detector. Fortunately, in this case the event rate can be normalized to the 
corresponding case in charged current neutrino scattering, which produces a 
neutral pion in conjunction with the charged lepton.  Although this may dominantly 
occur due to intermediate state processes, such as production of a $\Delta$ baryon 
or N$^{*}$ followed by its decay back to a nucleon and a pion (see 
\cite{SinglePiProdNuReact} and reference therein), concern also arises regarding other 
processes, including those that may be coherent over the entire nuclear target with 
attendant amplification of the rate~\cite{EvidenceNuCoherentPiProd,CoherentPiProdNuReactions}. 
The approach taken above can also be applied to such calculations where again, the 
pion is not produced in an $s$-channel fashion from a nucleon or one of its $N^{*}$ 
or $\Delta$ excited states, but from a $t$-channel process, corresponding to the photon 
production above. We will discuss that process in a separate paper. 

Finally, we comment on the effect of replacing the $\omega$ by a $\rho^{0}$: For the 
case of $\rho$ meson exchange (in place of the $\omega$), the analogous plots  to 
those displayed for $\omega$-exchange are very similar in shape and absolute magnitude.
The ratio of separate contributions depends upon both the ratios of the vector meson-pion-photon 
coupling constants, $g_{V-\gamma-\pi}$, estimated in the Appendix, as well as the vector 
meson-$Z$ coupling constants, $g_{V Z}$.  We estimate $\sigma^\rho/\sigma^\omega \approx 
(g_{\rho-\gamma-\pi}/g_{\omega-\gamma-\pi})^2 \times (g_{\rho Z}/g_{\omega Z})^2 \approx 
1.6$, using parameter values taken from Ref.(\cite{PDG06}). The suppression arises primarily 
from the coupling associated with the meson$-\pi-\gamma$ vertex while the enhancement is 
due to $Z$-meson mixing.  See the Appendix \ref{sec:SubsidiaryIssues} for more details.  

The amplitudes for these processes are comparable, and significant interference is expected 
to be possible.  From this effect the overall cross sections may be modified by a factor between 
$0.07$ and $5.1$ for total destructive and constructive interference respectively.  The cross 
section yielded by the lower limit is well below current or expected future experimental 
sensitivities and therefore forms a negligible background in that case. These considerations 
are interesting for future work, because of the parallel process of coherent pion production: 
If the interference is destructive, there can be a significant difference between coherent pion 
production between charged and neutral current cases as this interference cannot occur in 
the charged current case. 

In conclusion, we remark that whether the interference is constructive or destructive, our 
results apply equally to neutrino and to antineutrino neutral current scattering, since that 
only affects the overall sign of the amplitude by the neutrino coupling to the $Z$-boson. 
Thus, the process discussed here must contribute equally to neutrino and to antineutrino 
neutral current scattering, and with the same energy and angular dependence for the 
appearance of the photon. 

\acknowledgments{We thank G. Garvey, W. Louis and G. Mills for discussions regarding 
the LSND and MiniBooNE experiments and their data. This work was carried out in part 
under the auspices of the National Nuclear Security Administration of the U.S. Department 
of Energy at Los Alamos National Laboratory under Contract No. DE-AC52-06NA25396.

After this work was completed, we learned of  a complementary analysis \cite{Hill:2009ek} 
that uses effective field theory techniques, rather than the phenomenological approach 
taken here.  The results are similar in character. 

\bibliographystyle{apsrev}
\bibliography{MyReferences}

\appendix 

\section{Subsidiary Issues}\label{sec:SubsidiaryIssues}

For completeness, we show the explicit calculations of the decay rate for the 
$\omega$ and the $Z-\omega$ mixing amplitude. 

\subsection{$\Gamma(\omega \rightarrow \pi + \gamma)$}\label{subsec:omegaDecay}

We calculate the $\omega$ decay width $\Gamma(\omega \rightarrow \pi + \gamma)$ 
from the phenomenological vertex ${\cal L_{I}} = eg_{\omega\gamma\pi} \epsilon_{\mu\nu\xi \sigma} 
\omega^{\mu}\partial^{\nu}\pi^{0}F^{\xi \sigma}$.  This same form arises in the triangle 
anomaly considered in H3.  Neglecting the $\pi^0$ mass, the squared amplitude of this 
process is
\begin{equation}
  \mathcal{A}^2 = -\frac{2e^2g_{\omega\gamma\pi}^2}{3}k \cdot q = 
  \frac{e^2g_{\omega\gamma\pi}^2 M_\omega^4}{6}, 
\end{equation}
where $k$ and $q$ are the photon and pion momenta, respectively.  Evaluating the decay width yields
\begin{equation}\label{eq:GammaOmegaPigamma}
 \Gamma(\omega \rightarrow \pi + \gamma) = \frac{\alpha g_{\omega\gamma\pi}^2 M_\omega^3}{24}.
\end{equation}
Fitting this to the observed decay width~\cite{PDG06}, we extract the coupling constant 
$g_{\omega\gamma\pi}~=~1.8/M_\omega$, which we use in the numerical examples throughout our 
analysis.  Evaluating the tree diagram without the Lorentz structure of the anomaly yields 
$g_{\omega\gamma\pi}~\sim~1.2/M_\omega$, i.e., a factor of $2/3$ smaller than is obtained from 
Eq.~(\ref{eq:GammaOmegaPigamma}).  These are written in the form of a dimensionless 
coupling relative to the $\omega$ mass to help illustrate the scale associated with the 
effective vertex and strong coupling strength.  A similar exercise may be performed with 
the $\rho^0$ decay, in which case one extracts $g_{\rho\gamma\pi}~=~0.55/M_\rho$.   

\subsection{$p^{2}$ dependence of the $Z-\omega$ mixing amplitude}\label{subsec:XomegaMixing} 

\begin{figure}
\includegraphics[scale=0.75]{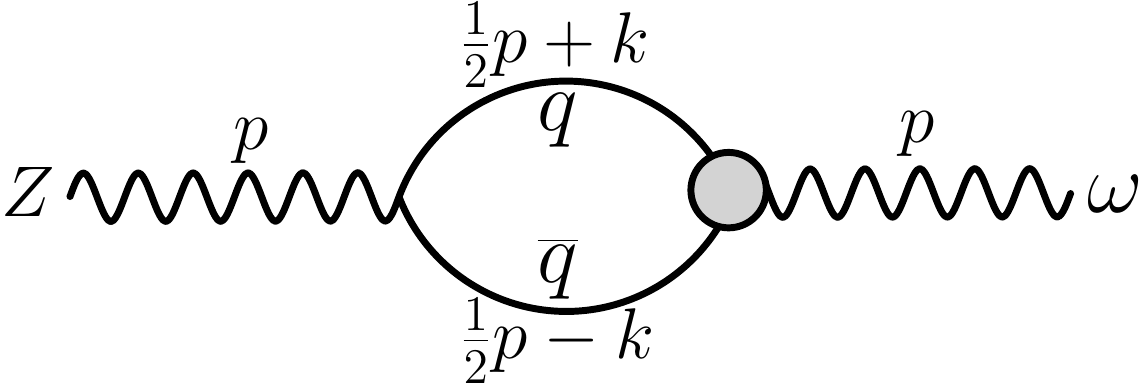}
\caption{Self energy diagram leading to $Z-\omega$ mixing.}
\label{fig:ZOmegaMixing}
\end{figure}

The $p^2$ dependence of the $Z-\omega$ mixing coefficient $g_{\omega Z}(p^2)$ may be 
found from the self-energy diagram shown in Fig.\ref{fig:ZOmegaMixing}.  Following 
references (\cite{GHT1,GHT2}), we parameterize the $\omega-q\bar{q}$ vertex by the form 
factor $g_{\omega q \bar{q}}M^2/(M^2-p^2)$, where the $g_{\omega q \bar{q}}~\approx~3.1$ 
coupling is extracted from $\omega~\rightarrow~\pi^0~\pi^+~\pi^-$ decay measurements~\cite{PDG06}, 
under the usual quark model assumption that inclusive processes may be well-approximated 
with the use of quark-hadron duality. The scale, $M$, defines the finite substructure of the 
$\omega$ meson.  Evaluating this diagram, we find the self-energy 
\begin{eqnarray}
 \Pi_{\mu\nu} &=& \frac{-8ig g_{\omega q\bar{q}}M^2\sin^2\theta_W}{3(2\pi)^d
 \cos\theta_W}\int_0^1 dx dy dz \delta(x+y+z-1) \\ \nonumber
 &\times& \int d^d\ell \frac{\left(1-\frac{2}{d}\right)\ell^2\eta_{\mu\nu} + z(z-1)\left(p^2
 \eta_{\mu\nu} - 2p_\mu p_\nu\right) + m_q^2\eta_{\mu\nu}}{\left(\ell^2 - p^2z(z-1) - 
 m_q^2(y+z) - M^2x\right)^3},
\end{eqnarray}
 where the d-dimensional integral is written to aid in the renormalization of the 
 divergent $\ell^2$ term via dimensional regularization and $m_q$ is the average light quark 
mass.  Considering only the terms that contribute to the $p^2$ dependence of $\omega-Z$ 
mixing and dropping logarithmic contributions, we find
\begin{equation}
 g_{\omega Z}(p^2) = \frac{-g g_{\omega q\bar{q}}M^2\sin^2\theta_W}{12\pi^2
 \cos\theta_W}\int_0^1 dz \int_0^{1-z} dx \frac{p^2z(z-1) + m_q^2}{p^2z(z-1) + 
 m_q^2+x(M^2-m_q^2)}.
\end{equation}
Taking the reasonable limit $m_q^2 \ll M^2,~p^2$, we evaluate the above integral to yield
\begin{eqnarray}\label{eq:gwz}
 g_{\omega Z}(p^2) &\approx& \frac{g g_{\omega q\bar{q}}M^2\sin^2\theta_W}
 {72\pi^2\cos\theta_W}\frac{1}{\eta^2} 
\left\{ 2\eta(1-\eta) + (2-3\eta)\ln\left(1-\eta\right) + \eta^3\ln\left(\frac{1-\eta}
{-\eta}\right) \right. \\ \nonumber 
&&~~~+ \left. \epsilon \left(\eta(4-\eta) + (4-3\eta)\ln(1-\eta) - \eta^2(6+\eta)
\ln\left(\frac{1-\eta}{-\eta}\right) \right) + \mathcal{O}(\epsilon^2)
\right\},
\end{eqnarray}
with the dimensionless quantities, $\eta~=~p^2/M^2$ and $\epsilon~=~m_q^2/M^2$.  
For the $p^2~<~0$ range relevant to this analysis, $g_{\omega Z}(p^2)$ is real.  The 
$\epsilon$ correction term is well-behaved and negligible for realistic parameter 
values where $\epsilon \sim 10^{-5}$.  The leading order term of Eq.~(\ref{eq:gwz}) 
breaks down for $-p^2 < 1~{\rm MeV}^2$ or $-\eta < 1.6\times 10^{-6}$.  The limiting 
behavior of $g_{\omega Z}(p^2)$ is easy to extract and is useful in estimating the 
physical parameter range
\begin{equation}
 g_{\omega Z}(p^2) = \frac{g g_{\omega q\bar{q}}M^2\sin^2\theta_W}{24\pi^2
 \cos\theta_W}\left\{ 
\begin{array}{ccc}
  -\left(1 + \frac{\ln (-\eta)}{\eta}\right) ~~~+ & \mathcal{O}\left(\frac{1}{\eta}\right) 
  ~~~~~& (-\eta \gg 1) \\
  \frac{1}{3}(1-\eta \ln (-\eta)) ~~+ & \mathcal{O}(\eta) ~~~~~& (-\eta \ll 1)
\end{array}.
 \right.
\end{equation}

For the neutrino beam energies studied here between $200-1000~\rm{MeV}$, we find that 
the average coupling $\bar{g}_{\omega Z}~=~600~{\rm MeV}^2$, under the assumptions 
that  $m_q~\sim ~3~\rm{MeV}$~\cite{PDG06} and $M~\sim~M_\omega$. 

\subsection{Relative strengths of $Z-\rho$ and $Z-\omega$ mixing}\label{subsec:relsize} 

On the basis of $SU(3)$ flavor symmetry, one expects a similar result  
for the $\rho^{0}$, except for the effect of isospin. In Fig.(\ref{fig:ZOmegaMixing}),  
the $\omega$ couples equally to the u and d quarks that contribute to the loop, which  
weights the $Zq\bar{q}$ couplings, viz.
\begin{eqnarray}
g_{Zu\bar{u}} &=& \frac{g}{4\cos\theta_W} (\frac{8}{3}\sin^2\theta_W-1)  \\
g_{Zd\bar{d}} &=& \frac{g}{4\cos\theta_W} (1-\frac{4}{3}\sin^2\theta_W) 
\end{eqnarray}
equally, whereas the $\rho^{0}$ does so with opposite signs due to isospin.  (The $\gamma_{5}$
parts of the couplings are omitted as irrelevant; they cannot contribute to the mixing since  there is 
only one momentum available to combine with the 4-dimensional Levi-Civita tensor that  comes 
from the fermion trace.) As might be expected from the fact that the $Z$-boson is dominantly isospin 
1 like the $\rho$, the $Z-\rho$ mixing is enhanced relative to the $Z-\omega$ mixing.

Combining these considerations with the slight phenomenological deviation from $SU(3)$  
flavor symmetry yields
\begin{eqnarray}
\frac{g_{\rho Z}}{g_{\omega Z}} &=& \frac{g_{Z d\bar{d}} - g_{Z u\bar{u}}}{g_{Z d\bar{d}} +
 g_{Z u\bar{u}}} \times \frac{g_{\rho q\bar{q}}}{g_{\omega q\bar{q}}} \\
 &\sim&  3\frac{1-2\sin^2\theta_W}{2\sin^2\theta_W}\times \sqrt{\frac{\Gamma(\rho\rightarrow\pi\pi)}
{ \Gamma(\omega\rightarrow\pi\pi\pi)}  
 \frac{ \phi(\omega\rightarrow\pi\pi\pi)}{\phi(\rho\rightarrow\pi\pi)}} \nonumber \\
 &=& 4.1, \nonumber
 \end{eqnarray}
where $\phi$ denotes the phase space integral for the decay, which suppresses the ratio by 
approximately $\sqrt{4\pi}$.  We take the numerical value of $\sin^2\theta_W  \sim 0.2396$ 
from the Particle Data Group extrapolation to small four-momentum transfer~\cite{PDG06}.  
The effect of this enhancement is included in our final discussion. 

\end{document}